\documentclass[aps,prd,twocolumn,a4paper]{revtex4-1}
\usepackage{ulem}
\usepackage{color}
\usepackage{graphicx}
\usepackage{epsfig}
\newcommand{\be}{\begin{equation}}
\newcommand{\ee}{\end{equation}}
\newcommand{\ba}{\begin{eqnarray}}
\newcommand{\ea}{\end{eqnarray}}
\newcommand{\nn}{\nonumber\\}
\begin{document}
\begin{flushright}
TIFR/TH/11-06
\end{flushright}
\title{A quasi-particle description of  $(2+1)$- flavor lattice QCD equation of state}
\author{Vinod Chandra$^{a}$}
\email{vinodc@theory.tifr.res.in}
\author{V. Ravishankar$^{b}$}
\email{vravi@iitk.ac.in}
\affiliation{$^{a}$ Department of Theoretical Physics, Tata Institute of Fundamental Research,
Homi Bhabha Road, Mumbai-400005, India.}
\affiliation{$^{b}$ Department of Physics, Indian Institute of
Technology Kanpur, Kanpur- 208 016, India.}
\date{\today}
\begin{abstract}
A quasi-particle model has been employed to describe the $(2+1)$-flavor lattice QCD 
equation of state with physical quark masses. The interaction part of the equation of state
has been mapped to the effective fugacities of otherwise non-interacting quasi-gluons and quasi-quarks.
The mapping is found to be exact for the equation of state. The model leads
 to non-trivial dispersion relations for quasi-partons. 
The dispersion relations, effective quasi-particle number densities, and trace anomaly have been 
investigated employing the model.
A Virial expansion for the EOS has further been obtained to
investigate the role of interactions in quark-gluon plasma (QGP). 
Finally, Debye screening in QGP has been studied employing the model.

\vspace{2mm}

\vspace{2mm}
\end{abstract}
\maketitle

\section{Introduction}
The main purpose of this article is 
to explore the quasi-particle picture of 
Quantum-Chromodynamics (QCD) at high temperature.
In particular, we wish to describe recent lattice data on (2+1)-flavor
QCD equation of state (EOS)by  employing 
a quasi-particle model~\cite{chandra1,chandra2,chandra4}. The EOS is an important quantity to 
study the properties of hot QCD matter which is commonly 
known as quark-gluon plasma (QGP) in 
relativistic heavy ion collisions at BNL, RHIC and
CERN, LHC. This study is mandated by the 
strongly interacting nature of QGP 
which has been inferred from the recent 
experimental observations at RHIC~\cite{flow,phen,phob,bram}. 
This observation is consistent  with the
lattice simulations of the EOS~\cite{boyd,karsch,cheng,cheng1,baz,fodor,fodor_new}, 
which predict a strongly interacting behavior even at temperatures which are a few $T_c$
(QCD transition temperature).

The most striking features of RHIC results~\cite{flow} are 
the large collective flow and strong jet quenching 
of high transverse momentum jets shown by QGP.
Similar conclusions have been drawn from the very recent preliminary results for 
Pb-Pb collisions at LHC at  $\sqrt{s}=2.76$ TeV~\cite{alice,alice1,hirano,hirano1}.
This has led to a tiny value of the shear viscosity to entropy density ratio
for QGP and near perfect fluid picture of 
QGP~\cite{visco1,visco2,visco3,visco4,visco41,visco5,muller,chandra3,chandra4}(except near the QCD 
transition temperature where the bulk viscosity of QGP is equally important as 
shear viscosity~\cite{khaz,mayer,buch}). 
In an attempt to appreciate this  result,
interesting analogies have been drawn with ADS/CFT
correspondence~\cite{dtson} as well as with some strongly coupled
classical systems~\cite{shuryak}. In any case, the emergence of strongly interacting behavior
puts into doubt the reliability of a large body of analyses which are based on ideal
or nearly ideal behavior of QGP in heavy ion collisions.

In the light of these observations, 
it would be right to state that QGP may 
lie in the strongly interacting domain (non-perturbative)
 of QCD. Therefore, lattice gauge theory~\cite{lat,gavai,kar1} 
would be the best approach to address the physics of QGP in RHIC
in terms of a reliable EOS which is very precisely 
evaluated~\cite{boyd,karsch,cheng,baz}. The lattice EOS is far from being ideal. 
The EOS is $\approx 10\%$ away from its ideal behavior even at $4 T_c$. 
However, in several works devoted to QGP ideal EOS is employed. 
This is certainly not desirable for QGP in RHIC.
Therefore, there is an urgent need to address this issue by developing models
to employ realistic QGP equations of state to investigate the bulk and 
transport properties of QGP. This could be done by casting hot QCD medium effects in terms of effective quasi-particle degrees of freedom.

There have been several attempts to describe QCD medium effects at high temperature in 
terms of quasi-particle degrees of freedom. These attempts include, (i) the effective mass approaches to study QCD thermodynamics
~\cite{peshier1,peshier2,pesh,alton,alton1}, and (ii) approaches based on the Polyakov 
loop~\cite{1,2,3,4,5}. A different approach in terms of quasi-particles, inspired by Landau theory of Fermi liquids has been proposed recently both for EOS based on pQCD~\cite{chandra1,chandra2,chandra3,chandra_nucla} and pure lattice guage theory~\cite{chandra4}.
This model is fundamentally different from the above two approaches, and quite powerful: 
it reproduces the EOS with remarkable accuracy, especially in the case of lattice EOS--where 
it is exact; the collective nature of the quasi-gluons is manifest in the single particle
dispersion relations. It is also successful in terms of predictions regarding the bulk and transport properties of QGP~\cite{chandra1,chandra2,chandra3,chandra4}.
Refs.~\cite{chandra3,chandra4} showed that the 
shear viscosity and its ratio with the entropy density ($\eta$, $\eta/s$) are highly sensitive to the interactions.
They could be taken of as good diagnostics to distinguish various EOS at RHIC. 

The model was tested only against the pure $SU(3)$ gauge theory EOS~\cite{chandra4}, where the 
description was, as we mentioned, exact. It was not 
employed for extracting a quasi-particle description in the case of 
full QCD, by the inclusion of quark sector. In this paper, we remedy this draw back and extend the model to the matter sector, by taking up a recently 
computed $(2+1)$-flavor lattice QCD equation of state with physical quark masses~\cite{cheng}. 
It is to be noted 
that this EOS has further been refined by improving the accuracy in~\cite{baz,cheng1}. 
Here, we adopt the philosophy same as in~\cite{chandra4}. Again, we map  $(2+1)$-flavor lattice QCD data for the EOS~\cite{cheng} in terms of quasi-particle degrees of freedom which are free up to effective gluon-fugacity, $z_g$ and effective quark fugacity, $z_q$. In this model, the strange quark sector is 
different from light quark sector due to contributions coming from the strange-quark mass. 
They are otherwise characterize by the same effective fugacity $z_q$.
Such a characterization could only be possible because the mass corrections from light quark-sector in the deconfined phase
of QCD are very very small. So, the model will be more realistic at higher temperatures.

Here, it is worth mentioning that such a quasi-particle description 
of the lattice QCD EOS could be thought of as a 
first step towards an effective field theory/ effective kinetic theory to 
explore complicated nature 
of strong interaction in QGP. Leaving these ambitious
investigations for future studies, here, we have  attempted to 
understand the role of QCD interactions in terms of a Virial expansion
for QGP employing the quasi-particle 
model. The Virial expansion has been obtained in terms of 
effective quasi-particle number densities which are ,in turn, expressed 
in terms of $z_{g/q}$. As we shall see that the Virial expansion 
of the EOS is very helpful to understand the 
role of strong interaction in QGP and may perhaps play crucial role 
in developing the effective models.

The paper is organized as follows. In section II, we introduce the quasi-particle  model and study its
features and physical significance. Here, we discuss the viability of the model by studying the 
temperature dependence of the quasi-particle pressure, and 
trace anomaly in terms of effective fugacities. We find that the model 
reproduces the EOS almost exactly. In section III, we discuss the 
physical significance and viability of the quasi-particle model.
In section IV, we discuss the implications of the model. Here, we 
propose a Virial expansion for QCD at high temperature, in terms of effective
quasi-particle number densities and explore the role of interaction in 
hot QCD. We further study the Debye screening and charge renormalization in hot QCD.
In section V, we present the conclusions and the future direction of the work.  

\section{The quasi-particle model}
Before, we introduce the quasi-particle description, let us define the notations. 
The quantities, $z_g$, and $z_q$ will denote effective gluon and quark/anti-quark 
fugacities respectively. The quasi-gluon equilibrium distribution function will be denoted by 
$f^{g}_{eq}$, quasi-quark/anti-quark
distribution function by $f^{q}_{eq}$ for light quarks ($u$,$d$), and $f^{s}_{eq}$
 for strange quark. The respective dispersions (single quasi-particle energy)
 will be denoted as $\omega^{g}_p$, $\omega^{q}_p$ and $\omega^{s}_p$. ${n_{g,q,s}}$ 
denotes the effective quasi-particle number densities. In all the physical
 quantities that will be discussed in the paper, the subscript $g$ will 
denote the gluonic contribution while $q$  and $s$ denote the contributions from the light quark sector and strange quark sector respectively. 

\subsection{Underlying distribution functions and effective fugacity}
We initiate the model with the ansatz that
the Lattice QCD EOS can be interpreted in terms of non-interacting
 quasi-partons having effective fugacities which encodes all the interaction effects. 
In the present case, we have three sector, {\it viz.}, the effective gluonic sector,
 the light quark sector, and the strange quark sector.
Here, the effective gluon sector refers to the contribution of 
gluonic action to the pressure which also involves contributions 
from internal fermion lines. Due to purely phenomenological reason,
 this sector can be recasted in terms of effective gluon quasi-particles 
(which are free gluons with effective fugacity). Similary the other two sectors 
also involve interactions among quark, anti-quarks, as well as  
their interactions with gluons. 
The effective gluon fugacity, $z_g$ is introduced to capture 
the interaction in the effective gluonic sector. 
On the other hand, $z_q$ captures interactions in other two sectors. The
ansatz can be translated to the form of the equilibrium distribution functions, 
$f^{g}_{eq}$, $f^{q}_{eq}$, and $f^{s}_{eq}$ as follows,

\ba
\label{eq1}
f^{g}_{eq} &=& \frac{z_g\exp(-\beta p)}{\bigg(1-z_g\exp(-\beta p)\bigg)},\nn
f^{q}_{eq} &=& \frac{z_q\exp(-\beta p)}{\bigg(1+z_q\exp(-\beta p)\bigg)},\nn
f^{s}_{eq}&=& \frac{z_q\exp(-\beta \sqrt{p^2+m^2})}{\bigg(1+z_q\exp(-\beta \sqrt{p^2+m^2})\bigg)},
\ea
where $m$ denotes the mass of the strange quark, which we choose to be $0.1 GeV$. $\beta=T^{-1}$ denotes inverse of the 
temperature. Note that we are working in the units where Boltzmann constant, $K_B=1$, $c=1$, and $h/2\pi=1$. 
We use the notation $\nu_g=2(N_c^2-1)$ for gluonic degrees of freedom , $\nu_{q}=2\times 2\times N_c\times 2$ for light quarks, $\nu_s=2\times 2 \times N_c \times 1$ for the strange quark for $SU(N_c)$. Here, we are dealing with $SU(3)$, so  $N_c=3$. Since the model is valid in the deconfined phase of QCD (beyond $T_c$, $T_c$ is the QCD transition temperature),
the masses of the light quarks can be neglected. Therefore, in our model we only consider the mass for the strange quarks.  

As it is well known that QCD thermodynamics at high temperature 
is described in terms of a Grand canonical ensemble.
Now, it is straight forward to write down an effective 
Grand canonical partition function for hot QCD which yields 
the forms of the distribution function given in Eq.(\ref{eq1}).
We denote the partition function by ${\bf Z}=(Z_g \times Z_q\times Z_s)$. 
The corresponding expressions in terms of $z_g$ and $z_q$ are as follows,
\ba
\label{leos}
\ln(Z_g)&=&-\nu_g V \int \frac{d^3 p}{8\pi^3} \ln(1-z_g\exp(-\beta p))\\
\ln(Z_q)&=&\nu_q V \int \frac{d^3 p}{8\pi^3} \ln(1+z_q\exp(-\beta p))\\
\ln(Z_s)&=&\nu_s V \int \frac{d^3 p}{8\pi^3} \ln(1+z_q\exp(-\beta (\sqrt{p^2+m^2}))\\
\ln ({\bf Z})&=&\ln(Z_g)+\ln(Z_q)+\ln(Z_s).
\ea
Now using the well known thermodynamic relation,
$P\beta V=\ln(Z)$, we can match the {\it rhs} of Eq.(5), with the lattice 
data for the pressure for $(2+1)$- flavor QCD~\cite{cheng}, where $P$ denotes the pressure and V denotes the volume.
From this relation, we can in principle determine 
the temperature dependence of $z_g$ and $z_q$. As emphasized earlier, $z_g$
is determined from the contribution to the lattice pressure purely from gluonic action. This 
particular contribution to the pressure is denoted as $P_g$. Remaining part of the pressure
is utilized to fix the temperature dependence of $z_q$. Now, we have two relations and two unknowns.
Next, we discuss the determination of $z_g$ and $z_q$.

\subsubsection{Determination of $z_g$ and $z_q$}
We determine $z_g$ and $z_q$ numerically. 
$z_g$ has been determined using the relation, 
\ba
\label{ep}
P_g &=&
\frac{-\beta^{-4} \nu_g}{2\pi^2}\int_0^\infty du\ u^2\ \ln(1-z_g\exp(-u)).
\ea
 On the other hand, $z_q$ has been determined numerically using the following relation,
\ba
\label{eqg}
(P-P_g) &=&
\frac{\beta^{-4}}{2 \pi^2}\int_0^\infty du\ u^2\bigg\lbrace \nu_q \ln(1+z_q\exp(-u))\nn&&+
\nu_s \ln(1+z_q\exp(-\sqrt{u^2+\tilde{m}^2}))\bigg\rbrace,\nn
\ea
where $\tilde{m}=\beta m$ and $u$ ($u=\beta p$) is a dimensionless quantity. We have recorded those values of $z_g$ and $z_q$ which satisfy Eqs. (\ref{ep}) and
(\ref{eqg}). Next, we discuss their behavior with temperature.
 
\subsubsection{Behavior of $z_g$ and $z_q$}
The determination of the the quasi-parton distribution functions
given in Eq.({\ref{eq1}) is complete once the temperature dependence of 
$z_g$ and $z_q$ is fixed. The behavior of $z_g$ and $z_q$ as a function of temperature is shown in Fig. 1 and  in Fig. 2 respectively.
Clearly, both of them acquire their ideal values (unity) only asymptotically.
At lower temperatures, the magnitude of both $z_g$ and $z_q$ is smaller indicating the 
larger strength of interactions there. 

\begin{figure}
\vspace{2mm}
\includegraphics[scale=.40]{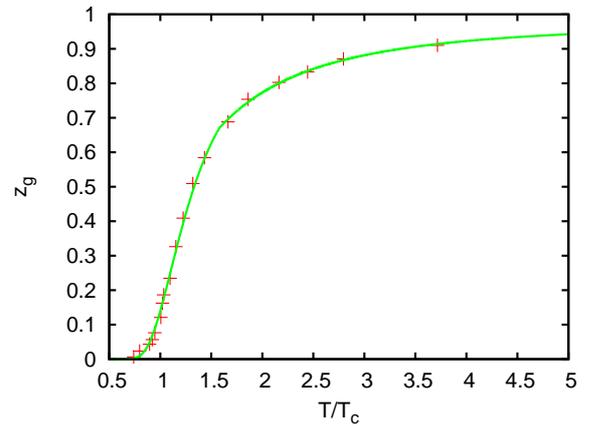} 
\caption{(Color online)  Behavior of $z_g$ as a function of $T/T_c$.
The points denote the values obtained from lattice data and solid line denote the 
fitting function. The fitting parameters are listed in Table. I.}
\vspace{2mm}
\end{figure}

\begin{figure}
\vspace{2mm}
\includegraphics[scale=.40]{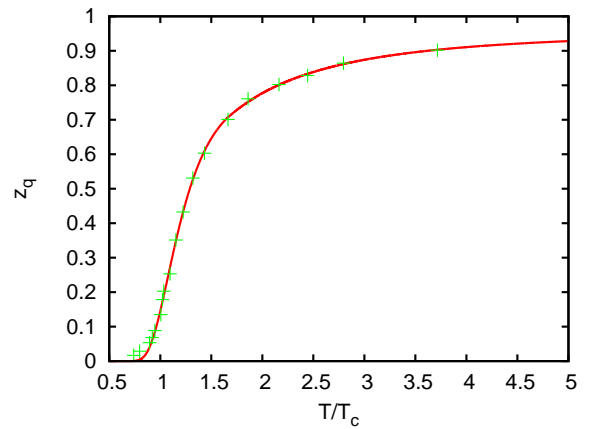} 
\caption{(Color online)  Behavior of $z_q$ as a function of $T/T_c$.
The points denote the values obtained from lattice data and solid line denote the 
fitting function. The fitting parameters are listed in Table. I.}
\vspace{2mm}
\end{figure}

For further analysis, we seek analytic forms for $z_g$ and $z_q$
as a function of temperature, which would render the computation 
more amenable. At this juncture, we note that there are limited number of 
lattice data for the pressure for a huge range of temperature ($(0.5-4.0) T_c$).
Since the effective fugacities have been determined from 
the lattice pressure,  the limitations get inherited by them as well, which
further passed on to other thermodynamic quantities such as 
energy density, entropy density and trace anomaly.
We hope that the functional form is not drastically altered by future refinements in lattice data.

We find that there is no universal functional form that 
describes the data in the full range of 
temperatures, either in the gluonic sector or the quark sector.
There are some common features though.
Both the sectors are characterized by a 'low temperature' and a 'high temperature' regime, with the cross over temperatures given by 
$x_{g,q}\equiv T_{g,q}/T_c\sim 1.68, 1.70$ respectively. The functional forms on 
either side of $x_{g,q}$ are the same for both the sectors, but with different parameters. Thus, when $x < x_{q,g}$, the good fitting function has the form $a_{g,q}\ \exp(-b_{g,q}/x^5)$. In the complimentary case $x> x_{q,g}$, it has the form $a^\prime_{g,q}\ \exp(-b^\prime_{g,q}/x^2)$. The latter form is mandated by the fact that at high temperature 
the trace anomaly, $\frac{\epsilon-3 P}{T^4}$ predominantly goes as $1/T^2$~\cite{cheng} in lattice QCD. 
We list the fitting parameters for $z_g$ and $z_q$
in Table I. 

\begin{table}
\label{table1}
\caption{Fitting parameters for $z_g$ and $z_q$} 
\begin{tabular}{|l|l|l|l|l|}
\hline
$z_{g,q}$& $a_{g,q}$ &$b_{g,q}$&$a^\prime_{g,q}$&$b^\prime_{g,q}$\\
\hline\hline
Gluon &0.803$\pm$0.009&1.837$\pm$0.039&0.978$\pm$ 0.007& 0.942$\pm$0.035\\ 
Quark &0.810$\pm$ 0.010& 1.721$\pm$0.040&0.960  $\pm$ 0.007&
 0.846 $\pm$  0.033
\\
\hline
\end{tabular}
\end{table}

We shall utilize these forms to study temperature dependence of the trace anomaly in later part of the paper.
We shall see that these forms correctly reproduce the high and low temperature behavior of the trace 
anomaly.

 Temperature dependences of effective fugacities, in Fig. 1 and Fig. 2 reveal
that the effective gluon and quark fugacities are of same order of magnitude for the whole range of temperature. 
This indicates that effective gluons and quarks contribute equally in our description. 
Its possible physical consequences,  and an understanding from basic calculations
 in QCD is beyond the scope of the present work and will be a matter of future investigations.
 
It is worth noting that  effective fugacity
descriptions have been earlier employed in condensed
matter systems in the last decade. To study the nature of Bose-Einstein (BE)
condensation transition in interacting Bose gases, a parametric
EOS in terms of the effective fugacity has been proposed
by Li {\it et. al}~\cite{li}. This provides a scheme to explore
the quantum-statistical nature of the BEC transition.
There have been other works to study the non-interacting
BE systems in harmonic trap~\cite{25} as well 
interacting bosonic systems~\cite{26}.
Moreover, effective fugacity description has been used
for a unitary fermion gas by Chen {\it et al}~\cite{24} for studying
thermodynamics with non-Gaussian correlations.None of them
employed the effective dispersion relations which we obtain naturally in this
work. We shall now proceed to discuss the physical significance of
 the quasi-particle model and its viability.

\section{Physical significance and viability of the model}
\subsection{The modified dispersion relations}
It has been emphasized in Ref.~\cite{chandra2} that the 
physical significance of effective fugacity could be seen in terms of 
modified dispersion relations. The effective fugacities modify the single quasi-parton energy
as follows,
\ba
\label{eq13}
\omega^g_p&=&p+T^2\partial_T ln(z_g)\nn
\omega^q_p&=&p+T^2\partial_T ln(z_q)\nn
\omega^s_p&=&\sqrt{p^2+m^2}+T^2\partial_T ln(z_q).
\ea
These dispersion relations can be interpreted as follows.
The single quasi-parton energy not only depends upon the momentum
but also gets contribution from the collective excitations of the quasi-partons.
The second terms is like the gap in the energy due to the presence of 
quasi-particle excitations. This immediately reminds us of Landau's theory of Fermi -liquids.
Therefore, it is safe to say that the present quasi-particle model is in the 
spirit of Landau theory of Fermi liquids. These modified dispersion relations in Eq.(\ref{eq13}) 
have emerged from the thermodynamic definition of the average energy of the system, due to the 
temperature dependent fugacities, $z_{g/q}$. Let us consider the expression for the 
energy-density, ${\epsilon }$ obtained in terms of the Grand Canonical partition function, ${\bf Z}$ as,
\be
\epsilon=-\frac{1}{V}\frac{\partial \ln({\bf Z})}{\partial\beta}. 
\ee
Substituting for the effective partition function (${\bf Z}=Z_{g}\times Z_{q}\times Z_s$), we 
obtain,
\ba
\epsilon&&\equiv \frac{\nu_g}{8\pi^3}\int d^3 p \bigg(p+T^2\partial_T \ln(z_g))\bigg)f^{g}_{eq}\nn
&&+\frac{1}{8\pi^3}\int d^3 p \bigg\lbrace\bigg(p+T^2\partial_T \ln(z_q))\bigg)\nu_q f^{q}_{eq}\nn&&+
\bigg(\sqrt{p^2+m^2}+T^2\partial_T \ln(z_q)\bigg)\nu_s f^{s}_{eq})\bigg\rbrace\nn
&&\equiv 3 (P_g+P_q+P_s)+\frac{T^2\partial_T \ln(z_g)}{2\pi^2}\int d^3p\ \nu_g f^g_{eq}\nn
&&+\frac{T^2\partial_T \ln(z_q)}{2\pi^2}\int d^3p\ \bigg(\nu_q f^q_{eq}+\nu_s f^s_{eq}\bigg).
\ea
The above equation can be recasted employing the expression for the pressure in 
terms of the temperature dependent, $z_{g/q}$ as, 
\be
\frac{(\epsilon-3 P)}{T^4}=T \frac{\partial}{\partial T}\bigg(\frac{P}{T^4}\bigg)
\ee

Therefore, these modified dispersion relations
naturally ensure the thermodynamic consistency condition in high temperature QCD, and lead to the 
trace anomaly, which we have discussed, in detail, in the next subsection.
Moreover, these effective fugacities can be expressed in terms of effective quasi-particle
number densities. These number densities leads to a simple Virial expansion for the EOS which we 
shall discuss in the next section.  

Next, we look at the group velocity of quasi-partons.
The group velocity can be obtained as $\vec{v}_{p}=\partial_{\vec{p}} \omega_p$.
It is easy to see that the modified term in the dispersion relations is purely temperature
dependent, therefore it will not change the group velocity of a quasi-parton.
$\vec{v}_p=\hat{p}$ for quasi-gluons and quasi-quarks $(u,d)$ and $\vec{v}_p=\frac{\vec{p}}{\sqrt{p^2+m^2}}$.
for strange quarks. The dispersion relation in Eq.(\ref{eq13}) contributes to the trace anomaly in 
hot QCD which we shall discuss soon.

\begin{figure}
\vspace{2mm}
\includegraphics[scale=.40]{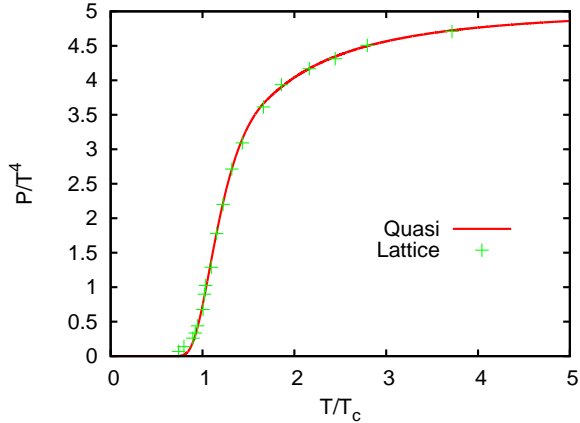} 
\caption{\label{feos} (Color online)  Behavior of $P/T^4$ as a function of $T/T_c$.
The quasi-particle pressure is obtained by using the fitting parameters for $z_g$ and $z_q$
listed in Table I. Lattice Data are also shown as points.}
\vspace{2mm}
\end{figure}

Let us now discuss the significance of  of gluon condensate,
in the hot QCD thermodynamics. In this context, D'Elia, Giacomo, and Meggiolaro~\cite{ella} 
have studied the electric and magnetic contributions to the condensate, and shown that
near $T_c$ the former vanishes, however the latter remains unchanged. 
These authors investigated such effects by analyzing the two-point correlation functions both in pure-gauge sector, and the full QCD~\cite{ella,ella1}. It has been shown in~\cite{casto} that the effects of the gluon condensate are significant for $T\geq T_c$, and becomes vanishingly small beyond $2 T_c$. Therefore, in the effective mass description of hot QCD for the temperatures, $T=1-2\ T_c$, one needs to consider the contributions of the condensate, while comparing the predictions on thermodynamic observables with the lattice QCD data.
However, in our model, $z_g$, $z_q$ capture these effects, and we do not need to incorporate the contributions 
separately. This can be understood as follows, 
In the lattice data employed here, normalization of the pressure and energy density 
were chosen such that at $T=0$, these quantities vanish~\cite{cheng}, and the effects of the gluon condensate
may be significant at higher temperatures. These effects are well captured in the trace anomaly in lattice QCD, 
which is the basic quantity computed in the lattice. And, all other thermodynamic quantities have been derived from the trace anomaly. In turn, these effects are automatically encoded in the pressure, the energy-density {\it etc.}. In our study, such effects have been captured in the effective fugacities, $z_{g/q}$ from the beginning, since, 
we have determined them from the lattice data on pressure. The gluon condensate contribute significantly to the energy density, entropy density, and the trace anomaly through the temperature derivatives of the $z_{g/q}$, in terms of modified dispersion relations.

\begin{figure}
\vspace{2mm}
\includegraphics[scale=.40]{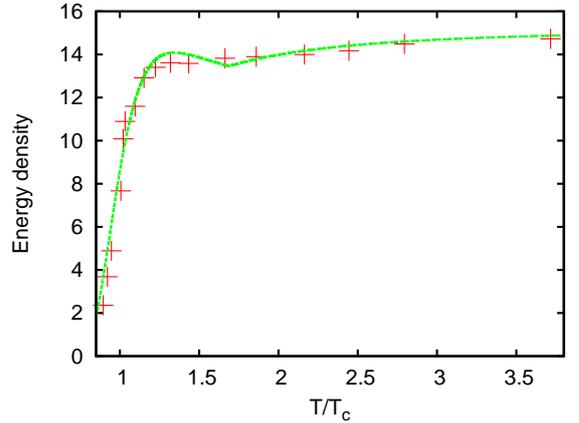} 
\caption{(Color online)  Behavior of $\epsilon/T^4$ as a function of $T/T_c$
in the quasi-particle model. The lattice results are shown as points and the 
solid line shows the quasi-particle result obtained by utilizing the fitting parameters,  
$a_{g,q},b_{g,q}$, and $a^\prime_{g,q},b^\prime_{g,q}$ listed in Table. I.}
\vspace{2mm}
\end{figure}

\subsection{Viability of the model}
As it has been already emphasized in the previous section that the model yields
(2+1)-flavor lattice QCD EOS almost perfectly. 
To check further the viability of the model, we study the 
temperature dependence of the quasi-particle pressure,energy density, and the 
trace anomaly and  check them against the direct lattice results. 
Lets us first discuss the temperature dependence of the 
pressure. We have plotted the quasi-particle pressure 
along with the lattice data in Fig. 3. We find that the agreement between 
the lattice data and quasi-particle model for the EOS is almost perfect beyond $T_c$.
The temperature dependence of the energy density as a function of temperature 
is shown in Fig. 4. The quasi-particle results agree well with the lattice data beyond $T_c$.

\begin{figure}
\vspace{2mm}
\includegraphics[scale=.40]{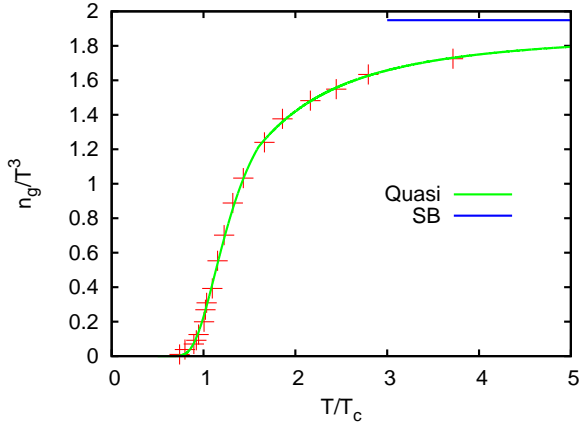} 
\caption{\label{fng} (Color online)  Behavior of $n_g$ as a function of $T/T_c$.
$N_g$ is obtained by employing the discrete data points for $z_g$ as well as fitting parameter for
of $z_g$ listed in Table. I. The Stefan-Boltzmann (SB) limit of $n_g$ is also shown.}
\vspace{2mm}
\end{figure}

\begin{figure*}
\vspace{2mm}
\includegraphics[scale=.40]{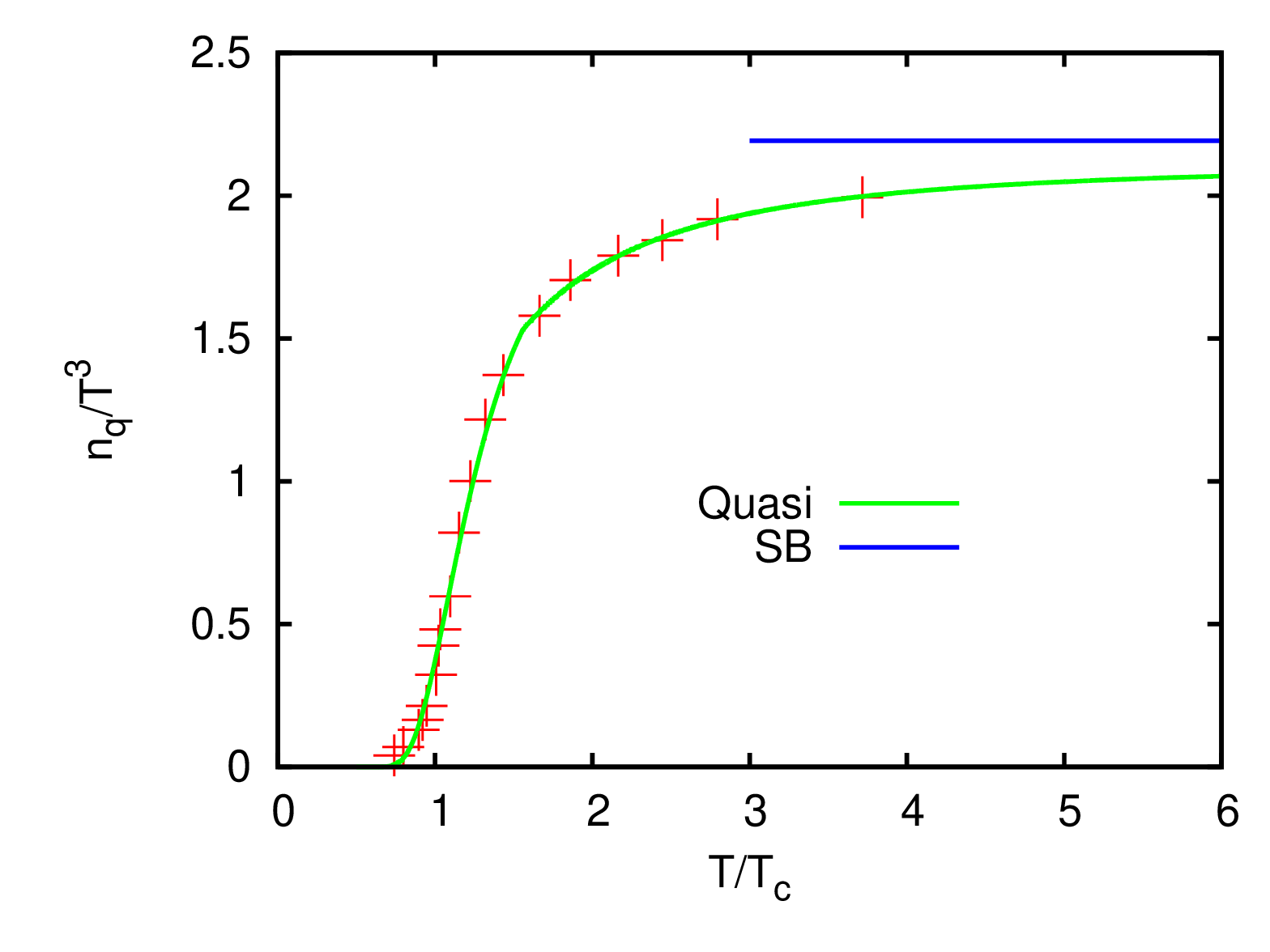}
\hspace{-.4cm}
\includegraphics[scale=.40]{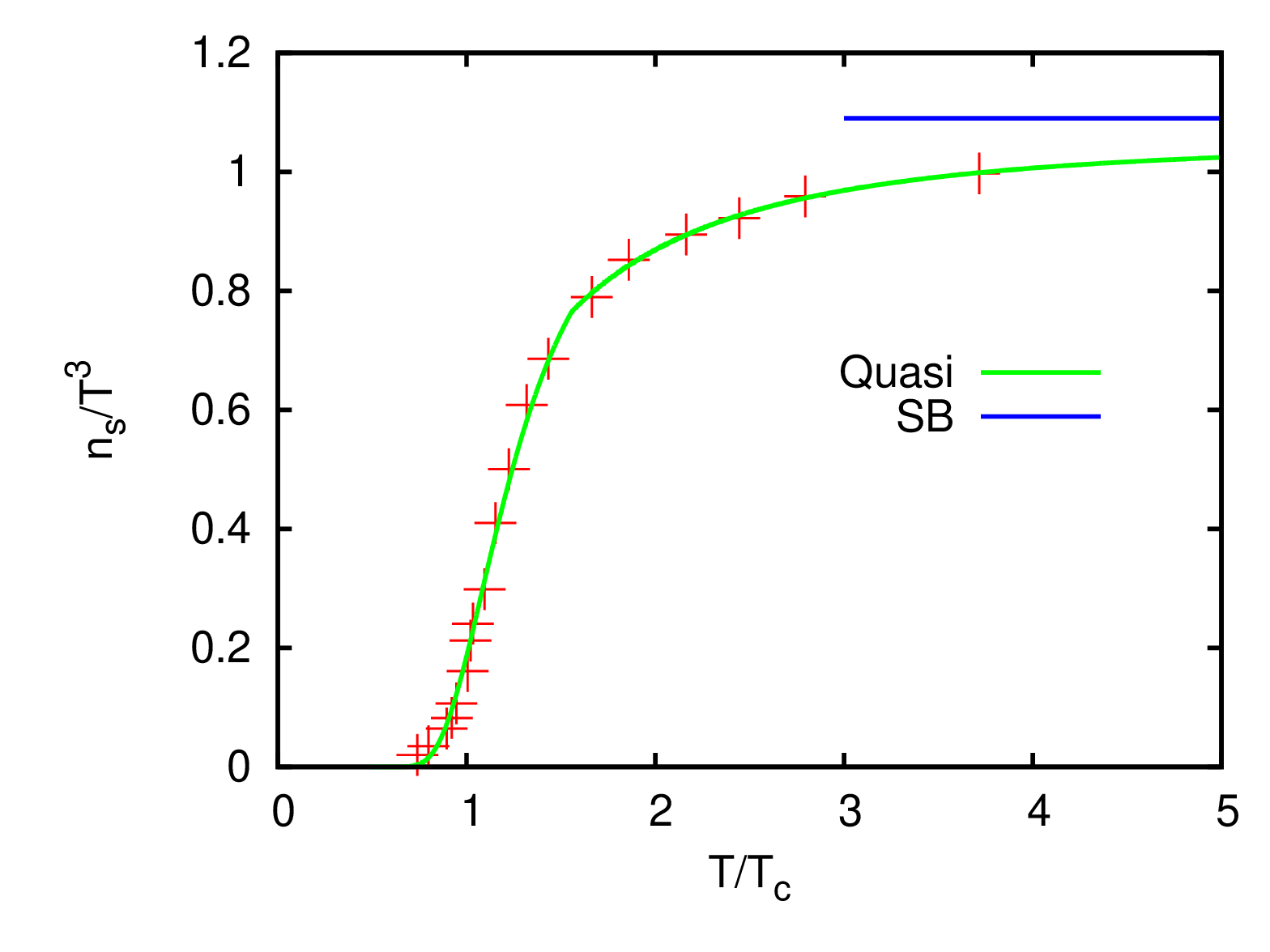}
\caption{\label{fnq} (Color online)  Behavior of $n_q$ (left) and $n_s$ (right) as a function of $T/T_c$.
They are obtained by employing the discrete data points for $z_q$ as well as fitting parameter for
of $z_q$ listed in Table. I including the strange quark mass dependence. The Stefan-Boltzmann (SB) value is also shown 
beyond $3 T_c$.}
\vspace{2mm}
\end{figure*}

Encouraged from the crucial observation that lattice and quasi-particle model predictions 
are the same, we now proceed to study the trace anomaly as a function of temperature 
obtained from the quasi-particle model.

\subsubsection{The trace anomaly}
Trace anomaly gets contribution from all 
the three sectors. The gluonic and light quarks contributions 
come purely from the modified part of the dispersion relations. On the 
other hand, in the strange quark sector trace anomaly gets additional contribution from the 
mass. We denote the trace anomaly by $\Delta= (\epsilon-3P)\equiv \Delta_g+\Delta_q+\Delta_s$.
\ba
\label{tr}
\frac{\Delta_g}{T^4}&=& T\partial_T \ln(z_g) \frac{{n}_g}{T^3}\nn
\frac{\Delta_q}{T^4}&=& T\partial_T \ln(z_q) \frac{{n}_q}{T^3}\nn
\frac{\Delta_s}{T^4}&=& T\partial_T \ln(z_g) \frac{{n}_s}{T^3}\nn
\ea
where $ n_g$, $ n_q$ and $n_s$ are the effective number densities for the 
quasi-partons and are defined by,
\ba
\label{effn1}
{n}_g&=&\frac{\nu_g}{2\pi^2} \int^\infty_0 dp p^2 f^g_{eq}\equiv T^3 \frac{\nu_g PolyLog[3,z_g]}{\pi^2}\nn
{n}_q&=&\frac{\nu_q}{2\pi^2} \int^\infty_0 dp p^2 f^q_{eq}\equiv T^3 \frac{-\nu_q PolyLog[3,-z_q]}{\pi^2}\nn
{n}_s&=&\frac{\nu_s}{2\pi^2} \int^\infty_0 dp p^2 f^s_{eq}\equiv T^3 \frac{-\nu_s PolyLog[3,-z_q]}{\pi^2}\nn
&& -\frac{3\tilde{m}^2}{\pi^2} \ln(1+z_q).\nn
\ea
We shall first discuss the behavior of these effective number densities as 
a function of temperature and thereby the temperature dependence of the trace anomaly
in hot QCD. We determine the effective number densities exactly by the numerical evolutions of the 
integrals in Eq.(\ref{effn1}). They are represented in terms PolyLog functions 
merely to understand the non-trivial temperature dependence of the 
effective number densities. 
Here, we see that $n_g/T^3$, and $n_{q/s}/T^3$ scales with $T/T_c$ in a non-trivial way. 
Their behaviors with temperature are shown in Figs. \ref{fng} and \ref{fnq}.

We have also shown the Stefan-Boltzmann value for the number densities in Figs. \ref{fng} and \ref{fnq}.
$({n_g/T^3})\vert_{SB}=\nu_g \zeta(3)/\pi^2$; $({n_q/T^3})\vert_{SB}=3\nu_q \zeta(3)/{4 \pi^2}$; and
$({\nu_s/T^3})\vert_{SB}= 3\nu_s \zeta(3)/{4 \pi^2}$. Note that the SB limit has not been obtained even at 
$T= 5 T_c$. The effective number densities are roughly $\sim  10\%$ away at this temperature in all 
the three sectors. This is just the reflection of the fact that lattice EOS itself is away from 
SB limit there.

Using Table. I, the quantities $T\partial_T \ln(z_g)$ and $T\partial \ln(z_q)$ can easily be obtained as,
\ba
T\partial_T \ln(z_g)=
\left\{ \begin{array}{rcl}
 \frac{5 b_{g}}{x^5}; &\mbox{x$\le$ $x_g$} &\\
\frac{2 b^\prime_g}{x^2};&\mbox{x$>$ $x_g$}&
\end{array} \right.
\nn
T\partial_T \ln(z_q)=
\left\{ \begin{array}{rcl}
 \frac{5 b_{q}}{x^5}; &\mbox{x$\le$ $x_q$} &\\
\frac{2 b^\prime_q}{x^2}; &\mbox{x$>$ $x_q$}.&
\end{array} \right.
\ea
\begin{figure}
\includegraphics[scale=.40]{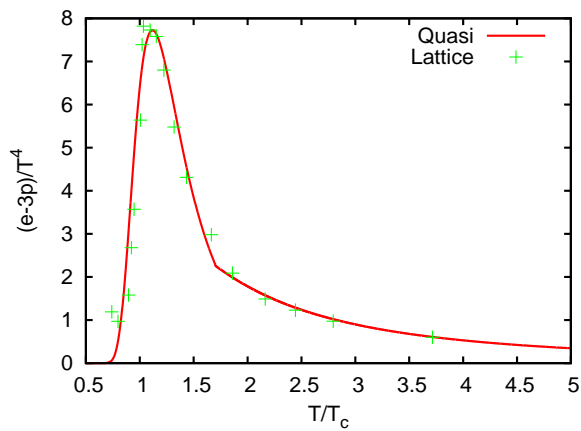} 
\caption{\label{tr_fit} (Color online)  Behavior of trace anomaly as a function of $T/T_c$.
The solid line denotes the values obtained from the quasi-particle description and 
the points denotes the lattice results.}
\end{figure}

It is now straight forward to compute the trace anomaly employing the fitting functions for 
the effective fugacities; and  effective quasi-particle number densities using Eq.(\ref{tr}).
Behavior of $\Delta$ along with corresponding lattice values has been shown in Fig. 7.
As it is clear from Fig. \ref{feos} and Fig. \ref{tr_fit},
pressure and trace anomaly computed by employing the 
quasi-particle model show good agreement with the 
lattice data of the same. This sets the utility of the model.
Once these two quantities are known, it is straight forward to
determine the  energy density ($e$) and the entropy density($s$) by the 
standard thermodynamic relations, $e=3 P_{quasi}+\Delta$, 
$s=\frac{e+P_{quasi}}{T}\equiv 4 P_{quasi}+\Delta$.

If we see, closely the  behavior of energy-density or the trace anomaly 
as a function of temperature in Figs. 4, and 7,  we observe that around $T=1.7 T_c$, both the quantities are 
not showing  smooth behavior. There is no physical reason associated with this. This is merely the artifact of the two 
distinct fitting functions for $z_g$ and $z_q$ below and above this temperature. At this temperature these functions 
have different slopes. This problem may not be present, if we could have found a single fitting functions 
for $z_g$ and $z_q$ for the whole range of temperatures. As emphasized earlier, these fitting functions were 
needed to compute, the temperature derivatives of $z_g$ and $z_q$. Moreover, these points merely 
reflect the fact that at this point, perhaps various thermodynamic quantities are changing their slopes, but smoothly,
which is not captured appropriately in the fitting functions. This may not be thought of as a serious problem 
since, one could compute the temperature derivatives of $z_g$ and $z_q$, directly by inverting relations 
in Eq. (12). In that case, we shall not get a continuous curve for energy density, rather discrete points and they
will perfectly match with the lattice predictions.

\section{Implications of the model}

\subsection{Virial expansion for hot QCD}
To translate QCD interactions in RHIC era in terms of a 
Virial expansion is quite a non-trivial task. There are 
very few attempts in this direction~\cite{cassing}.
Here, we see that the quasi-particle understanding of equation 
of state for QGP plays very crucial role to obtain a very simple
Virial expansion of the EOS in terms of effective quasi-particle number densities.
The model tells us that the Virial expansion in (2+1)-flavor QCD could be
subdivided in three sectors,{\it viz.}, gluonic, light-quark, and strange quark sector
and one can define the Virial expansion in each sector and finally combine them.

We begin with the purely gluonic sector first and
subsequently discuss the matter sector (light quarks and strange quarks).

\subsubsection{Virial expansion in purely gluonic sector}
To obtain the Virial expansion in this sector, we Taylor expand the pressure, 
$P_g$ and the effective gluon number density, ${n}_{g}$ in gluonic sector in the powers of $z_g$ (assuming $z_g<1$)
and eliminate the explicit dependence of 
$z_g$ from both the expressions. 
This technique which is the standard way to obtain the Virial expansion of a 
ideal Bose/Fermi gas with fixed number of particles~\cite{haung},
is equally applicable here, although the systems are physically distinct,
since $z_g$ and $z_q$ do not correspond to the 
particle conservation. The procedure is straight-forward. For the sake of completeness,
 we shall write a few steps.
The expressions for $P_g$ and $n_g$ in the powers of $z_g$ are obtained as,
\be
\label{pg}
\frac{P_g}{\nu_g T}=\frac{1}{\lambda_{th}^3}\sum_{l=1}^{\infty} b_{l}\ z_g^l.
\ee
\be
\label{ng}
\frac{n_g}{\nu_g}=\frac{1}{\lambda_{th}^3}\sum_{l=1}^{\infty} l \ b_{l} \ z_g^l.
\ee
The quantity $\lambda_{th}\equiv {1/T}$ is the thermal wave length of gluons and the 
coefficient $b_l$ is given as,
\be
b_l=\frac{1}{2\pi^2 \ l}\int_{0}^{\infty} u^2 \exp(-l u) du \equiv \frac{1}{\pi^2  l^4}. 
\ee

Consequently the ratio of Eq. ({\ref{pg}}) with Eq.({\ref{ng}}) and 
expanding it as,
\ba
\label{rt}
\frac{P_g}{{n}_g\ T}=\sum_{k=1}^\infty a_k (\tilde{N}_g \lambda_{th}^{3})^{k-1}
\equiv \frac{\sum_{l=1}^{\infty} b_{l}\ z_g^l}{\sum_{l=1}^{\infty} l \ b_{l} \ z_g^l},
\ea
where $\tilde{N}_g=\frac{n_g}{\nu_g}$ and $a_k$'s are the Virial coefficients.
Using, Eq. ({\ref{pg}})and Eq.({\ref{ng}}) in Eq.(\ref{rt}) and 
comparing the terms of order $z_g$, $z_g^2$ and $z_g^3$, we obtain,

\ba
\label{vircg}
a_1&=&1; a_2=-\frac{b_2}{b_1^2}\equiv-\frac{\pi^2}{2^4};\nn
a_3&=&\frac{\bigg(-2 b_3+\frac{4 b_2^2}{b_1}\bigg)}{b_1^3}\nn
a_4&=& \frac{1}{b_1^4} \bigg(-3 b_4 -\frac{20 b_2^3}{b_1^2}+\frac{18 b_2 b_3}{b_1}\bigg).
\ea
One can, in principle obtain all the Virial coefficients comparing various 
order coefficients of $z_g$ in Eq.({\ref{rt}}). 
Now, the Virial expansion up to $O((N^\prime_g \lambda_{th}^3)^3)$ can be written as,
\be
\label{vg}
\frac{P_g}{{n}_g\ T}= 1+a_2 (N^\prime_g \lambda_{th}^3) +a_3 (N^\prime_g \lambda_{th}^3)^2
+a_4 (N^\prime_g \lambda_{th}^3)^3.
\ee

If we exploit the temperature dependence of effective gluon number density shown in 
Fig. 5. We can compare the strength of various order terms in Eq.(\ref{vg}).
We find that third term is $<<$ second term and fourth term is $<<$ third term.
Remember that the Virial coefficients in Eq.(\ref{vircg}) do not acquire any temperature 
dependence and are same as those for an ideal gluonic plasma with temperature independent 
fugacity. The interactions merely renormalizes the number density of quasi-gluons.
This confirms our view point that hot QCD medium effects can  entirely be mapped in to the 
non-interacting/weakly interacting quasi-particle degrees of freedom.
The validity of the Virial expansion in this sector is ensured by the fact that 
$N^\prime_g \lambda_{th}^3<< 1$.

Let us now move to the matter sector and first discuss the light quark sector followed by the 
strange quark sector.

\subsubsection{Virial expansion in the matter sector}
We shall exactly follow the same procedure discussed earlier to obtain 
the Virial expansion. We denote the light quarks contribution to pressure as 
$P_q$ and contribution from strange quarks to $P_s$. Expanding these quantities 
along with effective number densities ${n}_q$ and ${n}_s$ in the 
power of $z_q$ ( assuming $z_q<1$), we obtain the following expressions,
\ba
\label{pq}
\frac{P_q}{\nu_q T}&=&\frac{1}{\lambda_{th}^3}\sum_{l=1}^{\infty}(-1)^{l-1}\ b_{l}\ z_q^l\\
\frac{P_s}{\nu_s T}&=&\frac{1}{\lambda_{th}^3}\sum_{l=1}^{\infty}(-1)^{l-1}\ b^\prime_{l}\ z_q^l
\ea
\ba
\label{nq}
\frac{n_q}{\nu_q}&=&\frac{1}{\lambda_{th}^3}\sum_{l=1}^{\infty} (-1)^{l-1} l \ b_{l} \ z_q^l\\
\frac{n_s}{\nu_s}&=&\frac{1}{\lambda_{th}^3}\sum_{l=1}^{\infty} (-1)^{l-1} l \ b^\prime_{l} \ z_q^l.
\ea
Where the coefficients $b^\prime_l$ is defined as,

\ba
b^{\prime}_l&=&\frac{1}{2\pi^2 \ l} \int_{0}^{\infty} u^2 \exp(-l \sqrt{u^2+\tilde{m}^2}) du \nn
&=& b_{l}-\frac{\tilde{m}^2}{4\pi^2 l^3}
\ea
Repeating the analysis same as for gluons, and 
denoting the Virial coefficients in light quark sector as $a^{q}_l$ and
strange quark sector as $a^{s}_l$, the Virial expansion 
for $P_q$ and $P_s$ would have the following forms,
\ba
\label{vm}
\frac{P_q}{n_q T}=a^q_1+a^q_2 (N^\prime_q \lambda_{th}^3)+a^q_3(N^\prime_q \lambda_{th}^3)^2+
a^q_4(N^\prime_q \lambda_{th}^3)^3\nn
\frac{P_s}{n_s T}=a^s_1+a^s_2 (N^\prime_s \lambda_{th}^3)+a^s_3(N^\prime_s \lambda_{th}^3)^2+
a^s_4(N^\prime_s \lambda_{th}^3)^3.\nn
\ea
 where $N^\prime_q =\frac{n_q}{\nu_q}$ and $N^\prime_s =\frac{n_s}{\nu_s}$.
In principle, all the Virial coefficients ($a^q_l$ and $a^s_l$) are possible to compute 
in terms of $b_l$ and $b^\prime_l$. We shall only discuss up to the
fourth Virial coefficient. These coefficients are obtained as follows,
\ba
a^q_1&=&1;\ a^q_2=\frac{b_2}{b_1^2}=\frac{\pi^2}{2^4};\ a^q_3=\frac{\bigg(-2 b_3+\frac{4 b_2^2}{b_1}\bigg)}{b_1^3}\nn
a^q_4&=& \frac{1}{b_1^4} \bigg(3 b_4 +\frac{20 b_2^3}{b_1^2}-\frac{18 b_2 b_3}{b_1}\bigg);\nn
a^s_1&=&1;\ a^s_2=(b_2-\frac{\tilde{m}^2}{2^5\pi^2})/(b_1-\frac{\tilde{m}^2}{2^2\pi^2})^2;\nn
a^s_3&=& \frac{\bigg(-2 b^\prime_3+\frac{4 ({b^\prime_2})^2}{b^\prime_1}\bigg)}{({b^\prime_1})^3}; \nn
a^s_4&=& \frac{1}{({b^\prime_1})^4} \bigg(3 b^\prime_4 +\frac{20 {(b^\prime_2})^3}{({b^\prime_1})^2}-\frac{18 b^\prime_2 b^\prime_3}{b^\prime_1}\bigg).
\ea

Again the dominant contribution is from the second terms in the Virial expansion of 
$P_q$ and $P_s$ in Eq.(\ref{vm}).This we have observed by exploiting the 
temperature dependence of the effective quasi-particle number densities shown in 
Fig. 6.

Since the total quasi-particle pressure is $P=P_g+P_q+P_m$, Virial expansion of the 
full EOS can be obtained by using the individual Virial expansions obtained 
in the effective gluonic sector (EGS), and the matter sector ( Eqs.(\ref{vg}) and (\ref{vm}). 
Note that all the Virial coefficients in  EGS as well as in the 
matter sector are independent of temperature. The information about the 
interaction has been captured in the effective quasi-parton number-densities.
The second Virial coefficient is negative in the gluonic sector and
positive in the quark sector. This is expected from the quantum statistics of 
quasi-gluons and quasi-quarks. It is straightforward an exercise to determine the other thermodynamic observables in
term of effective quasi-parton number densities by using the well known thermodynamic relations.
These expressions will also contain the temperature derivatives of effective number densities in addition.
In other words, both $z_g$ and the modification factor to the dispersion relations, 
$T\partial_T \ln(z_g)$, $T\partial_T \ln(z_q)$  will appear in their expressions. 
Finally, the validity of the Virial expansion in the matter sector is ensured by the
fact that $N^\prime_{q,s} \lambda_{th}^3 << 1$.  

This is perhaps the first time, we have obtained such a simple 
Virial expansion for hot QCD where interactions appear as suppression factors
through the effective quasi-parton number densities. This has only been 
possible due to the quasi-particle description of hot QCD. Interesting enough, such a 
description works well down to temperatures which are of the order $1.0 T_c$.
The Virial expansion here highlights the role of interactions in hot QCD.
The Virial expansion may possibly play crucial role
to explore a quantitative understanding of  Fermi liquid like picture of hot QCD interactions as 
indicated by our quasi-particle description and also play important role to 
develop effective field theory and effective kinetic theory for such a quasi-particle model.
We shall leave these interesting issues for the future investigations.

At this juncture, we wish to mention that there has been a very
recent attempt~\cite{Virial} to study the nuclear matter EOS at sub-nuclear
density in a Virial expansion of a non-ideal gas. The Virial expansion is 
obtained by considering the fugacities for various species such as neutron, proton etc.
The method to obtain the Virial coefficients is standard one as employed in 
the present work. However, the major difference between the two is in the 
physical meaning of the fugacities.

\subsection{Comparison with other approaches}
We now intend to compare our quasi-particle model with other 
existing models. In the recent past~\cite{1,2,3,4} and in a very recent work~\cite{5}, effects of hot QCD
medium have been interpreted in terms of single particle states (effective
gluons/quark-anti-quarks) via the Polyakov loop. In these approaches, the
expectation value of the Polyakov loop appears in the effective gluon/quark-antiquark
distribution functions~\cite{4}. It provides a suppression factor in the form of a effective fugacity
to an isolated particle with color quantum numbers.
On the other hand, there have been successful attempts to 
encode the high temperature QCD medium effects in terms of 
effective thermal masses for quasi-partons~\cite{pesh,peshier1,peshier2,alton,alton1}.
In the recent past, effective mass models have been employed to describe $(2+1)$-flavor QCD~\cite{thaler,kalman} and the agreement was found to be good. The effective mass models, which we discussed so far are based on the lowest order results in perturbative QCD, equivalently leading order 
HTL results. These models are improved by incorporating the next order 
HTL contributions by Rebhan and Romatschle~\cite{rebhan}. Their predictions 
were shown to be in agreement with the lattice results including $(2+1)$-flavor QCD.
Furthermore, there are other approaches which also involves quasi-particle picture of hot QCD along with the contribution from the gluon condensate\cite{casto}. We shall compare our model with these approaches one by one.

Let us consider the Polyakov loop approach first.
There are certain similarities and a number of differences between our
model and this approach. The similarities are, (i) both the approaches lead
to an effective description of hot QCD in terms of free quasi-particles, (ii) the
expectation value of the Polyakov loop which plays the role of effective fugacity
as well as effective quasi-parton fugacities ($z_{g/q}$) in our model appear as
the suppression factors in the corresponding quasi-parton distribution functions,
(iii) in both the approaches the group velocity ($v_{gr} = \partial_{\vec{p}} \omega_p$), 
remains unchanged, (iv) both the models are quite successful in reproducing the lattice data
on thermodynamic observables (For more details on Polyakov loop method see ~\cite{3}. 
Our model yield lattice EOS for $SU(3)$ pure gauge theory almost perfectly which the deviations
which are one part in a million~\cite{chandra3}), and same is true for the (2+1)-flavor lattice EOS,
and (v) the effective gluon distribution
function in~\cite{4, 5} has a similar mathematical structure as our model.
In spite of these similarities, our model is fundamentally distinct from
this approach. Our model is purely phenomenological, and is more in the
spirit of Landau’s theory of Fermi liquids. We list below the major differences
between the two approaches.
\begin{itemize}
\item The expectation value of the Polyakov loop appearing in the single
particle distribution function does not change the dispersion relation for
quasi-partons. On the other hand, in our model, we obtain non-trivial
quasi-parton dispersion relations.
$(T^2\partial_{T} ln(z_{g/q})$.
\item The Polyakov loop (its phase) appears as an imaginary chemical potential
in the single particle distribution functions~\cite{4,5}. This is unlikely
to happen in our model. The effective fugacities in our model cannot
be interpreted as chemical potentials (real/imaginary) (since there is
no conservation of particle number). They are introduced merely to
capture all the interaction effects present in hot QCD medium. 
\item Employing our quasi-particle model, one can study the 
the bulk and transport properties of hot and dense matter (QGP) in RHIC.
These studies have been reported for pure $SU(3)$ gauge theory in~\cite{chandra3,chandra4}
and will be presented separately for full lattice gauge theory in the near future.
\end{itemize}

Let us now compare our model with the effective mass models.
In this approach, lattice QCD data for the EOS had been 
interpreted in terms of effective thermal gluon mass and effective thermal 
quark mass. The quasi-particle model proposed in this paper 
is completely distinct from this model. The major difference is in the 
philosophy itself. The effective fugacities are not the 
effective masses and they can be interpreted as effective mass 
in some limiting case ($p<< T^2\partial_T \ln(z_{g/q})$).
Moreover, our approach explores the Fermi liquid like 
picture of hot QCD. Another major difference in two of the 
approaches can be realized in terms of group velocity $v_{gr}$,
$v_{gr}$ in two approaches is not the same. In the effective mass 
approaches $v_{gr}$ depends on thermal mass parameter ($\vec{v}_{gr}=\vec{p}/\sqrt{p^2+m(T)^2}$).
We have obtained a Virial expansion for hot QCD in terms of the the 
quasi-partons. This has not been done employing either
of these two models.

Let us now discuss the quasi-particle models which incorporate the 
effects of gluon condensate explicitly in the analysis~\cite{casto,quasi_new2}.
In these studies, the importance of the gluon condensate is highlighted,
and its effect on the thermodynamic observables in hot QCD was studied in detail~\cite{casto}.
In particular, Castorina and Mannarelli~\cite{casto}, have analyzed the thermodynamic 
properties of hot QCD between $1-2\ T_c$ by explicitly incorporating the 
gluon condensate along with the gluon, and quark quasi-particles with thermal masses.
The results show excellent agreement with the lattice predictions, both in the 
pure glue sector and the full QCD sector. Apart from the differences in 
the dispersion relations, and the philosophy with our model, there has been a 
very crucial difference. As emphasized earlier, in our model the effect 
of gluon condensate has been incorporated from the beginning, and not treated 
explicitly as in~\cite{casto}. However, both the models are equally 
successful to describe the lattice QCD thermodynamics.

There is an alternate way to interpret the effective 
fugacities in terms of effective mass,  as follows.
Let us suppose, $z_{g/q}\equiv\exp(-m_{eff}\vert_{g/q}/T)$. 
The quantity, $m_{eff}$ can be thought of as $m_{eff} = g^\prime(T) T$,
where $g^\prime$ is an effective coupling. It is to be observed that $z_{g/q}$ are of the order of 
$0.15$ around $T_c$; it leads to an estimate for 
$g^{\prime}\sim 2.0$. This indicates the non-perturbative nature 
of hot QCD matter near $T_c$. Moreover, $g^\prime$ becomes less than one beyond $1.3 T_c$, and this 
observation is valid in both effective gluon and matter sector.

\subsection{Debye screening mass and charge renormalization}
To investigate how the partonic charges 
modify in the presence of hot QCD medium, we 
consider the expression for the Debye mass derived 
in semi-classical transport theory~\cite{manual} in terms of equilibrium 
parton distribution functions. The same expression was obtained from the
chromo-electric response functions of QGP~\cite{chandra2}.
The Debye mass in terms of the quasi-parton distribution 
functions, which are obtained from the (2+1)-lattice QCD EOS, is given by, 
\begin{equation}
\label{dby}
M^{2}_{D}=-2 N_c Q^2 \int  \frac{d^3p}{8\pi^3} \partial_{p} f^{g}_{eq}+ Q^2\int \frac{d^3 p}{8\pi^3} \partial_{p} (4 f^{q}_{eq}+2 f^{s}_{eq}),
\end{equation}
where $Q^2$ is the effective coupling which 
appears in the transport equation. If one assumes QGP as an ideal system of massless gluons and quarks,
Eq.(\ref{dby}) reproduces the leading order HTL result for the Debye mass; with the identification that $Q^2\equiv g^2(T)$ ($g(T)$ is QCD running coupling constant at finite temperature).

Employing the distribution functions displayed in  Eq.(\ref{eq1}) to Eq. (\ref{dby}), we obtain the following expressions for the 
\ba
\label{dby1}
M^2_D&=& Q^2 T^2 \bigg(\frac{N_c}{3}\frac{6\ Polylog[2,z_g]}{\pi^2}\nn&&+\frac{1}{2}\times \frac{-12\ PolyLog[2,-z_q]}{\pi^2}
-\frac{\tilde{m}^2}{4\pi^2}\ln(1+z_q)\bigg).\nn
\ea
While determining the Debye mass in Eq.(\ref{dby1}) from Eq.(\ref{dby}), we 
employ Eq.(\ref{eq1}) and the following standard integrals,
\ba
\int_{0}^\infty p^2\ dp \frac{z_{g/q} \exp(-\beta p)}{(1 \mp z_{g/q}\ \exp(-\beta p))^2} 
&&\equiv\pm \frac{2}{\beta^3} PolyLog[2,\pm z_{g/q}]\nn
\int_{0}^\infty p\ dp \frac{z_q \exp(-\beta p)}{(1 +z_q\ \exp(-\beta p))^2} &&\equiv \frac{1}{\beta^3}\ln(1+z_q).
\ea

The Debye mass with the Ideal EOS($z_g=1$,\ $z_q=1$) will be,
\be
\label{mdi}
(M^{I}_D)^2=Q^2 T^2 \bigg(\frac{N_c}{3}+\frac{1}{2}-\frac{\tilde{m}^2}{4\pi^2}\ln(2)\bigg).
\ee 

To analyze the role of interactions, we define the effective charges $Q_g$, $Q_q$, and $Q_s$
as,
\ba
Q_g&=& Q \ \bigg\lbrace \bigg(\frac{6 PolyLog[2,z_g]}{\pi^2}\bigg)^{1/2}\bigg \rbrace\nn
Q_q&=& Q \ \bigg\lbrace \bigg(\frac{-12 PolyLog[2,-z_q]}{\pi^2}\bigg)^{1/2}\bigg \rbrace\equiv Q_s\nn
\ea 

Debye mass could be written in terms of these effective charges as, 
\be
\label{mdni}
M^2_D=\bigg\lbrace \frac{N_c}{3}Q^2_g +\frac{1}{2}(Q^2_q + Q^2_s)\bigg\rbrace T^2-Q^2\ T^2 \frac{\tilde{m}^2}{4\pi^2}\ln(1+z_q).
\ee            
As stated earlier, in the ideal limit Eq.(\ref{mdni}) will reduce to Eq.(\ref{mdi}).
The quantities, $Q_g$, and $Q_{q,s}$ approach to $Q$.
This observation tells us that 
interactions merely renormalize the effective partonic charges. In fact, the effective charges are 
reduced as compared to $Q$ and asymptotically approach to the ideal value, $Q$. 

\begin{figure}
\vspace{2mm}
\includegraphics[scale=.40]{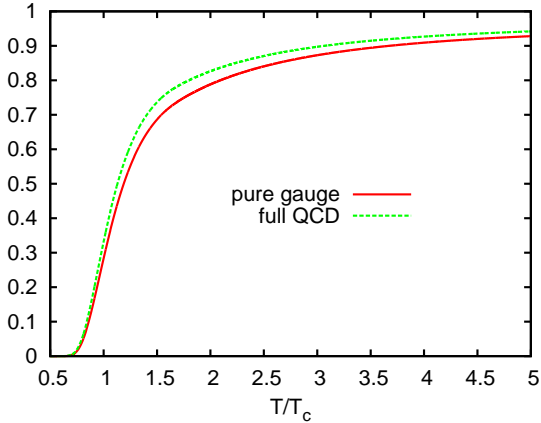} 
\caption{\label{dmass} (Color online) Behavior of $\mu_d$ as a function of $T/T_c$.
Note that $\mu_d$ approaches to the ideal value only asymptotically.}
\vspace{2mm}
\end{figure}
To see, how interactions modify Debye screening mass, we
consider the ratio $\mu_{d}={M_D}/{M^{I}_D}$. The behavior of$ \mu_d$ as a function of temperature is shown in Fig. \ref{dmass}. $\mu_d$ approaches the ideal value unity only asymptotically and $\mu_d\le 1$.
This implies that the presence of interactions suppresses the Debye mass as compared to its ideal 
counterpart.

Next, we compare $M_D$ with the Debye mass obtained in lattice QCD.
Let us first discuss the Debye screening mass computed in lattice gauge theory.
It has been calculated in pure gauge theory ($N_f = 0$)~\cite{kaz},
in 2-flavor QCD ($N_f = 2$)~\cite{kaz1,ma1}, and in $2+1$-flavor QCD~\cite{pet}.
The lattice data on Debye mass have been fitted with the simple ansatz motivated 
by leading order result on Debye mass, $m^{L}_D\equiv A\ m^{LO}_D$, where $m^{L}_D$ denotes the 
lattice data and, $m^{LO}_D\equiv \sqrt{(1+\frac{N_f}{6})}g(T) T$ denotes the leading order Debye mass. Here, 
$g(T )$ is the two loop running coupling constant. This form fits the data quite well if $A \approx 1.4-1.6$\cite{pet}.

In our case (see Eq.\ref{mdni}), $Q^2$ is a free parameter. We can fix it to match the Debye mass with the lattice 
result, $m^{L}_D$. This leads to,
\ba
 Q^2&=& A^2\frac{g^2(T)}{(6\ PolyLog[2,z_g])/{\pi^2}}, \nn
 Q^2&=& A^2\ g^2(T) \bigg\lbrace\bigg((6\ PolyLog[2,z_g]\nn
 &&-12\ PolyLog[2,-z_q])/{\pi^2}\bigg)
 -\frac{\tilde{m}^2}{4\pi^2}\ln(1+z_q)\bigg\rbrace^{-1} \nn
\ea
in EGS and full QCD with $N_f$ flavors respectively.
In other words, the Debye mass obtained from the quasi-particle model can exactly
be matched with the lattice results for the Debye mass. 
The Debye mass is needed to determine the transport parameters for quark-gluon plasma 
in RHIC~\cite{chandra3}.  It is of interest to 
derive the form of heavy quark potential~\cite{akhilesh,chn_2}  employing the formalism of 
chromo-electric response functions~\cite{chandra2,akhilesh}. While deriving the potential, 
one should keep in mind the fact that hadronic phase to quark-gluon plasma transition 
is a crossover~\cite{crs1,crs2} rather than a true phase transition.
These issues will be taken up in a separate communication in near future.

\section{Conclusions and future prospects}
In conclusion, a quasi-particle model for $(2+1)$-flavor lattice 
QCD has been proposed which is valid in the deconfined phase of QCD.
The interactions have been encoded in 
to the effective gluon and quark fugacities. These effective fugacities 
non-trivially modify the single quasi-parton energies and 
lead to the trace anomaly in hot QCD. The description accurately reproduce
the lattice QCD pressure, energy-density, and the trace anomaly. In particular, the model 
accurately reproduce their low and high temperature behavior. We find that
the model is fundamentally  distinct from the other
quasi-particle models (effective thermal mass, Polyakov loop models, and models with gluon condensate).

Employing the model, temperature dependence of the effective quasi-particle number densities 
has been obtained. A Virial expansion for QGP has been proposed
in terms of effective quasi-particle number densities.
The Virial expansion of the quasi-particle equation of state
gets contribution from three sectors, {\it viz.}, the effective gluonic sector,
the light quark-sector, and the strange quark sector. These sectors were dealt separately and
eventually lead to the complete Virial expansion. This is perhaps the first time such an Virial expansion 
has been proposed for hot QCD. Interestingly, the Virial expansions came out to be 
mathematically similar as that for an ideal system of gluons, light quarks and, strange quark with 
temperature dependent fugacities. The Virial expansion has ensured that the 
quasi-particle are non-interacting. The interactions merely modulate the 
quasi-particle number densities and modify the single quasi-particle energies
in a non-trivial way.

The Virial expansion may play important role to 
explore the Fermi liquid like picture of hot QCD in the matter sector and, 
in building  effective kinetic theory 
with the quasi-particle model. The Virial expansion has revealed that the interactions
appear to various observables determined by employing the quasi-particle description 
only in two ways, either through the effective fugacities (act as modulation factors) or through 
the modified dispersion relation. Finally, Debye mass has been obtained employing the
expression obtained from semi-classical transport theory and effective coupling 
has been determined in terms of effective fugacities. This observation will be required in 
determining the transport coefficients for quark-gluon plasma. We find the the Debye mass 
obtained from the quasi-particle model can exactly be matched with the lattice results.

The implications of model to study the transport coefficients (shear viscosity, bulk viscosity) will be taken up 
in the near future. It would also be of interest to extend the present model in the case of 
finite baryon density, and studying quark-number susceptibilities.
It is to be of great interest to establish possible connections of our 
quasi-particle model with the Polyakov loop models, which is a matter of 
future investigations.

\vspace{2mm}
\noindent{\bf Acknowledgements}: 
We are highly thankful to Prof. Saumen Datta for providing us the
lattice data, and Prof. Rajeev Bhalerao for many helpful discussions and help in the 
computational part of the work. VC is thankful to Prof. Edward Shuryak for many helpful 
suggestions and encouragement. We are indebted to the people of India for their 
invaluable support for the research in basic sciences in the country.

\end{document}